\documentclass[twocolumn,showpacs,preprintnumbers,prl,amsmath,amssymb]{revtex4}

\usepackage{graphicx}
\usepackage{dcolumn}
\usepackage{bm}
\usepackage{bbm}

\usepackage{hyperref}

\begin{document}
\title{Spin relaxometry of single nitrogen-vacancy defects in diamond nanocrystals\\
 for magnetic noise sensing}
\author{J.-P. Tetienne$^{1,2}$, T. Hingant$^{1,2}$, L. Rondin$^{1}$, A. Cavaill{\`e}s$^{1}$, L. Mayer$^{3}$, G. Dantelle$^{3}$, T. Gacoin$^{3}$, J. Wrachtrup$^{4}$, J.-F. Roch$^{2}$, V. Jacques$^{1,2}$}
\email{vjacques@ens-cachan.fr}
\affiliation{$^{1}$Laboratoire de Photonique Quantique et Mol\'eculaire, CNRS and ENS Cachan UMR 8537, 94235 Cachan, France \\
$^{2}$Laboratoire Aim\'e Cotton, CNRS, Universit\'e Paris-Sud and ENS Cachan, 91405 Orsay, France \\
$^{3}$Laboratoire de Physique de la Mati\`ere Condens\'ee, CNRS and Ecole Polytechnique UMR 7643, 91128 Palaiseau, France \\
$^{4}$3. Physikalisches Institut and SCoPE, Universit{\"a}t Stuttgart, 70550 Stuttgart, Germany}

\begin{abstract}

We report an experimental study of the longitudinal relaxation time ($T_1$) of the electron spin associated with single nitrogen-vacancy (NV) defects hosted in nanodiamonds (ND). We first show that $T_1$ decreases over three orders of magnitude when the ND size is reduced from 100 to 10 nm owing to the interaction of the NV electron spin with a bath of paramagnetic centers lying on the ND surface. We next tune the magnetic environment by decorating the ND surface with Gd$^{3+}$ ions and observe an efficient $T_{1}$-quenching, which demonstrates magnetic noise sensing with a single electron spin.  We estimate a sensitivity down to $\approx 14$ electron spins detected within $10$~s, using a single NV defect hosted in a $10$-nm-size ND. These results pave the way towards $T_1$-based nanoscale imaging of the spin density in biological samples.

\end{abstract}

\maketitle

The ability to detect spins is the cornerstone of magnetic resonance imaging (MRI), which is currently one of the most important tools in life science. However, the sensitivity of conventional MRI techniques is limited to large spin ensembles, which in turn restricts the spatial resolution at the micrometer scale~\cite{Lee2001,Ciobanu2002}. Extending MRI techniques at the nanoscale can be achieved at sub-Kelvin temperature with magnetic resonance force microscopy, through the detection of weak magnetic forces~\cite{Mamin2007,Degen2009}. Another strategy consists in directly sensing the magnetic field created by spin magnetic moments with a nanoscale magnetometer. In that context, the electron spin associated with a nitrogen-vacancy (NV) defect in diamond has been recently proposed as an ultrasensitive and atomic-sized magnetic field sensor~\cite{Taylor2008}. In the last years, many schemes based on dynamical decoupling pulse sequences have been devised for sensing ac or randomly fluctuating magnetic fields with a single NV spin~\cite{Maze2008,Hall2009,Laraoui2010,deLange2011}. These protocols recently enabled nuclear magnetic resonance measurements on a few cubic nanometers sample volume~\cite{Mamin2013,Staudacher2013} and the detection of a single electron spin under ambient conditions~\cite{Grinolds2013}. \\
\indent An alternative approach for sensing randomly fluctuating magnetic fields -- {\it i.e.} magnetic noise -- is based on the measurement of the longitudinal spin relaxation time ($T_1$) of the NV defect electron spin. Using an ensemble of NV defects and a $T_1$-based sensing scheme, Steinert {\it et al.} recently demonstrated magnetic noise sensing with a sensitivity down to $1000$ statistically polarized electron spins, as well as imaging of spin-labeled cellular structures with a diffraction-limited spatial resolution ($\approx 500$ nm)~\cite{Steinert2012}. Bringing the spatial resolution down to few nanometers could be achieved by using a single NV defect integrated in a scanning device, {\it e.g.} with a nanodiamond (ND) attached to the tip of an atomic force microscope (AFM)~\cite{Balasubramanian2008,Rondin2012}. With this application in mind, we study here the $T_1$ time of single NV defects hosted in NDs, as a function of ND size and magnetic environment. We first report a decrease of $T_1$ over three orders of magnitude when the ND size is reduced from 100 to 10 nm. This behavior is explained by considering the interaction of the NV spin with a bath of intrinsic paramagnetic centers lying on the ND surface. We next tune the magnetic environment by decorating the ND surface with paramagnetic molecules. As expected, a strong $T_1$-quenching is observed when the surface spin density is increased. From our data, we estimate a sensitivity of $T_1$-based relaxometry down to $\approx 14$ electron spins detected within $10$~s, using a single NV defect hosted in a $10$-nm ND.  \\
\begin{figure}[t]
\begin{center}
\includegraphics[width=0.42\textwidth]{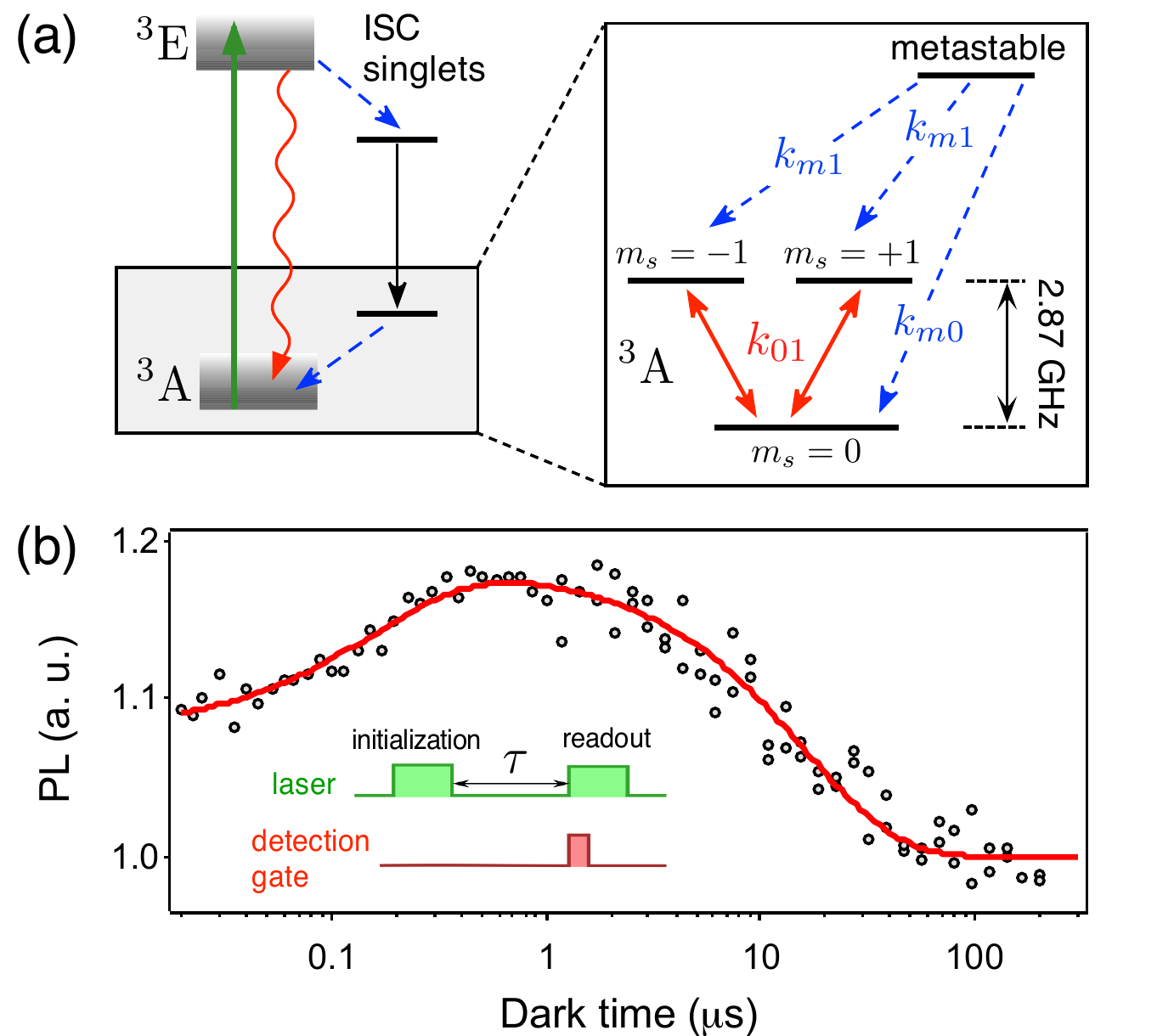}
\caption{(a)-Simplified energy-level structure of the NV defect. The zoom indicates the energy levels and transition rates used for studying the NV defect spin dynamics in the dark. (b)-Integrated PL signal as a function of the dark time measured for a single NV center hosted in a $20$-nm ND. The solid line is a fit to Eq. (\ref{eqPL}), which yields $T_1=16\pm 1$ $\mu$s and $T_m=160 \pm 30$ ns. The experimental sequence used to measure $T_1$ is shown in inset. Laser pulses with 3-$\mu$s duration are used both for initialization of the NV defect in $m_s=0$ and for spin-state read-out by recording the PL signal in a detection window corresponding to the first 300 ns of the optical pulses.} 
\label{Fig1}
\end{center}
\end{figure}
\indent The NV defect ground state is a spin triplet ($S=1$) with a zero-field splitting $D=2.87$ GHz between a singlet state $m_{s}=0$ and a doublet $m_{s}=\pm 1$ [Fig.~\ref{Fig1}(a)]. Owing to spin-dependent intersystem-crossing (ISC) towards intermediate singlet states, optical pumping leads to an efficient spin polarization into the $m_s=0$ spin sublevel, while the spin state can be readout through spin-dependent photoluminescence (PL)~\cite{Manson2006}. These two properties enable the measurement of the $T_1$ relaxation time of the NV defect electron spin by using the simple sequence depicted in the inset of Fig.~\ref{Fig1}(b). After initialization into the $m_{s}=0$ spin sublevel with an optical pulse, the NV defect is kept in the dark for a time $\tau$, causing the system to relax towards a mixture of states $m_s=0,\pm 1$. The resulting electron spin state is readout by applying a second optical pulse. For a sufficiently short integration time (300 ns in this work), the readout PL signal $\mathcal{I}(\tau)$ can be written as $\mathcal{I}(\tau)\approx A_0 n_0(\tau)+A_1 [n_{+1}(\tau)+n_{-1}(\tau)]$, where $A_0$ and $A_1<A_0$ are the PL rates associated with spin states $m_s=0$ and $m_s=\pm 1$, respectively, and $n_{0,\pm 1}(\tau)$ are the spin populations before applying the readout optical pulse. These populations are evaluated within the simplified four-level model shown in Fig.~\ref{Fig1}(a), which includes the ground state spin sublevels $m_s=0,\pm1$ and the lowest-lying singlet state, thereafter referred to as the metastable state. We define $T_1$ as the decay time of the population $n_0$, hence $1/T_1=3k_{01}$, where $k_{01}$ is the two-way transition rate between $m_{s}=0$ and $m_{s}=\pm 1$. At short time scale, the spin populations are also affected by relaxation from the metastable state which decays towards the ground state spin sublevels as $n_m(\tau)=n_m(0)e^{-\tau/T_m}$, where $T_m=(k_{m0}+2k_{m1})^{-1}$ is the metastable state decay time. The value of this parameter is $\approx200$~ns~\cite{Manson2006,Robledo2011}. Using classical rate equations within this four-level model, the PL signal $\mathcal{I}(\tau)$ can be written~\cite{SI}
\begin{equation} \label{eqPL}
\mathcal{I}(\tau)=\mathcal{I}(\infty)\left[1-C_m e^{-\tau/T_m}+C_1 e^{-\tau/T_1}\right] \ .
\end{equation}
The expressions of $\mathcal{I}(\infty)$, $C_m$ and $C_1$ are given in the Supplemental Material \cite{SI}. A typical measurement of $\mathcal{I}(\tau)$ is shown in Fig.~\ref{Fig1}(b) for a single NV defect hosted in a $20$-nm-size ND, together with a fit to Eq.~(\ref{eqPL}) yielding $T_1=16\pm 1 \ \mu$s. This value is almost two orders of magnitude smaller than the one measured for single NV defects hosted in bulk diamond samples~\cite{Jarmola2012}.\\
\indent To understand this behavior, the $T_1$ time was studied as a function of the ND size. We started from commercially available NDs (SYP $0.05$ and $0.25$, Van Moppes SA) produced by milling type-Ib high-pressure high-temperature (HPHT) diamond crystals with a high nitrogen content ([N]$\approx 200$~ppm). The formation of NV defects was carried out using high energy (13.6 MeV) electron irradiation followed by annealing at $800^{\circ}$C under vacuum. The irradiated NDs were then oxidized in air at $550^{\circ}$C during two hours in order to remove graphitic-related defects on the surface and produce stable NV defects~\cite{Rondin2010}. The NDs were finally spin cast on a glass cover slip and studied using a scanning confocal microscope combined with an AFM (Attocube Systems), all operating under ambient conditions~\cite{SI}. For each photoluminescent ND, the PL intensity autocorrelation function was first recorded in order to verify that a single NV defect was hosted by the crystal. For a set of single NV defects in isolated NDs, the $T_1$ time was measured by fitting the relaxation curve $\mathcal{I}(\tau)$ to Eq.~(\ref{eqPL}) and AFM measurements were used to infer the ND diameter $d_0$, defined as the maximum height in the AFM scan [Fig.~\ref{Fig2}(a)]. The relaxation rate $1/T_1$ is plotted as a function of the ND size in Figure~\ref{Fig2}(b) for a set of $51$ single NV defects in isolated NDs with $d_0$ ranging from $7$~nm to $88$~nm. An increase of the relaxation rate over three orders of magnitude is observed when the ND size decreases. Indeed, $T_1$ ranges from a few $\mu$s for the smallest ($<10$ nm) NDs to up to $1$~ms for the biggest ones ($>60$ nm) [Fig.~\ref{Fig2}(a)\&(b)].\\ 
\begin{figure*}[t]
\begin{center}
\includegraphics[width=0.91\textwidth]{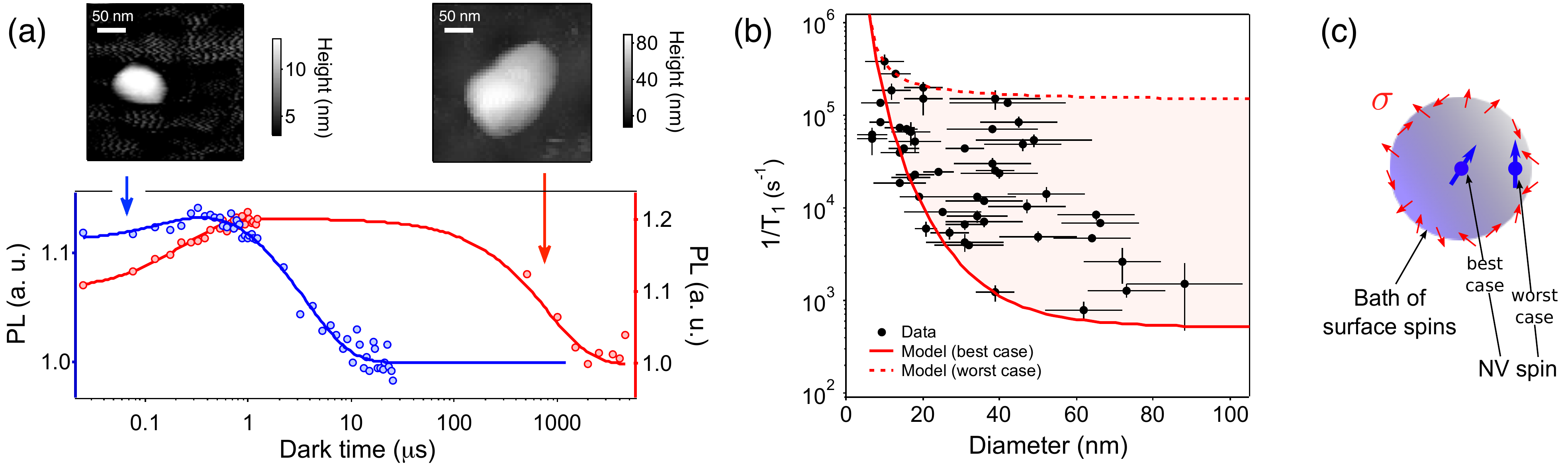}
\caption{(a)-Relaxation curves measured for a single NV defect hosted in a small ND ($d_0=13\pm3$ nm, blue curve) and a larger one ($d_0=73\pm10$ nm, red curve). Data fitting with Eq.~(\ref{eqPL}) (solid lines) gives $T_1$ values of  $3.6\pm 0.3$ $\mu$s and $802\pm 136$ $\mu$s, respectively. The corresponding AFM images are shown on top of the graph. (b)-Longitudinal spin relaxation rate $1/T_1$ of the NV defect electron spin as a function of the ND diameter. We note that no obvious correlations with the size was observed for the other fitting parameters in Eq.~(\ref{eqPL}), which were found to be $C_m=0.084\pm0.037$, $C_1=0.21\pm0.12$ and $T_m=198\pm72$ ns (mean $\pm$ s.d.). (c)-The ND is modeled as a sphere with diameter $d_{0}$ with a bath of randomly fluctuating surface spins with density $\sigma$. In (b), the markers are experimental data while the lines are the results of the calculation using this model for a NV spin located at the center of the sphere (solid line) and $3$ nm below the surface (dotted line). The parameters of the calculation are $ T_{1}^{\rm bulk}=2$ ms and $\sigma=1$ nm$^{-2}$~\cite{SI}.}
\label{Fig2}
\end{center}
\end{figure*}
\indent For NV defects hosted in bulk diamond samples, phonon-assisted processes are the main causes of longitudinal spin relaxation at room temperature, with $T_1$ lying in the 1-10 ms range~\cite{Jarmola2012}. Relaxation induced by paramagnetic impurities like nitrogen atoms (P1 centers), which are the most abundant paramagnetic defects in type-Ib diamond, dominates only at low temperature, and results in $T_1$ times that can be as long as $100$~s at 4 K \cite{Jarmola2012}. In NDs, a bath of paramagnetic centers covering the surface provides an additional channel for $T_1$ relaxation. These impurities have been identified by numerous studies~\cite{Shames2002,Dubois2009,Casabianca2011,Panich2012b,Tisler2009,Laraoui2012} and are mainly ascribed to dangling bonds with unpaired electron spins. For NDs with an oxygen-terminated surface, as those used in this work, Tisler {\it et al.} determined a density of surface spins $\sigma \approx 1-10$ spin/nm$^2$ using spin coherence measurements and ensemble EPR measurements~\cite{Tisler2009}. We attribute the shortening of $T_1$ of single NV spins in NDs to these surface paramagnetic centers (SPCs), which adds a contribution $k_{01}^{\rm spc}$ to the transition rate $k_{01}^{\rm bulk}$ of the bulk material, such that the overall rate is $k_{01}=k_{01}^{\rm bulk}+k_{01}^{\rm spc}$. \\
\indent This hypothesis is tested by modeling the ND as a sphere and the SPCs as an ensemble of randomly fluctuating spins with a surface density $\sigma$ [Fig.~\ref{Fig2}(c)]. The SPCs produce a fluctuating magnetic field ${\bf B}(t)$ with zero-mean $\langle B(t) \rangle=0$, that is characterized by the spectral densities $S_{B_k}(\omega)=\int_{-\infty}^{+\infty} B_k(t)B_k(t+\tau)e^{-i\omega\tau}{\rm d}\tau$, where the three components $k=x,y,z$ are assumed to be uncorrelated. For a central NV spin $S=1$ with an intrinsic quantization axis along $z$, one has \cite{Slichter}
\begin{equation} \label{eqT11}
k_{01}^{\rm spc}=\frac{\gamma_e^2}{2} \left[ S_{B_x}(\omega_0) + S_{B_y}(\omega_0) \right]   
\end{equation} 
where $\gamma_e$ is the electron gyromagnetic ratio and $\omega_0=2\pi D$ is the electron spin resonance (ESR) frequency of the NV defect. As highlighted by Eq. (\ref{eqT11}), longitudinal spin relaxation is caused by the transverse components of the magnetic noise at the ESR frequency of the central spin. Assuming correlation functions of the form $\langle B_k(0)B_k(\tau)\rangle=\langle B_k^2 \rangle e^{-|\tau|/\tau_c}$ where $\tau_c$ is the correlation time of the magnetic field and $\langle B_k^2 \rangle$ its variance, the relaxation rate reads
\begin{equation} \label{eqT12}
\frac{1}{T_{1}}=\frac{1}{T_{1}^{\rm bulk}} + 3\gamma_e^2 B_{\perp}^2 \frac{\tau_c}{1+\omega_0^2 \tau_c^2} \ .    
\end{equation} 
Here $ B_{\perp}^2 =\langle B_x^2 \rangle+\langle B_y^2 \rangle$ is the variance of the transverse magnetic field, and $1/T_{1}^{\rm bulk}=3k_{01}^{\rm bulk}$. Assuming $S=1/2$ surface spins, the variance $B_{\perp}^2$ is calculated by summing the dipolar field at the NV's location from each randomly oriented SPC and the correlation time $\tau_c$ is evaluated by considering intra-bath dipolar coupling~\cite{SI}. Since $B_{\perp}^2$ depends on the exact location and orientation of the NV defect inside the ND, the calculation is performed for two extreme configurations [Fig.~\ref{Fig2}(c)]. In the {\it best-case} scenario, the NV defect is located at the center of the sphere, while in the {\it worst-case} scenario, it is lying $3$~nm below the surface -- near the known photostability limit of the NV defect~\cite{Bradac2010} -- with its axis being parallel to the surface~\cite{SI}. As shown in Fig. \ref{Fig2}(b), the results of the model capture fairly well the experimental data with $\sigma=1$ nm$^{-2}$~\cite{Note}. More precisely the relaxation rate scales as $1/d_0^{4}$ which stems from the $1/d_0^{6}$ dependence of the spin-spin interaction integrated over a surface. This effect is responsible for the variation of $T_1$ over several orders of magnitude when the size of the ND decreases~\cite{SI}.\\
\indent In view of testing the ability of $T_1$ relaxometry to detect changes in the local magnetic environment, the ND surface was decorated with additional paramagnetic species. This was achieved by spin casting an aqueous solution of gadolinium perchlorate molecules Gd(ClO$_4$)$_3$, containing paramagnetic Gd$^{3+}$ ions ($S=7/2$), which is a well-known relaxation contrast agent in MRI. The $T_1$ time was measured for a set of 33 single NV defects in isolated NDs (i) before any treatment, (ii) after a first treatment with 1 mM of Gd$^{3+}$ solution, and (iii) after a second treatment with 10 mM. The substrate was patterned with a metallic grid for precise and repeatable identification of each individual ND over repeated treatment steps. The histograms of the measured $1/T_1$ rates are shown in Fig. \ref{Fig3}(a). The distribution is clearly shifted towards higher relaxation rate after each treatment step, which indicates that single NV defects feel the magnetic noise induced by the external Gd$^{3+}$ ions. In Fig.~\ref{Fig3}(b), the relaxation rate after the first and second treatment step is plotted as a function of the rate in the bare nanocrystal, {\it i.e.} before any treatment. Almost all the investigated NV defects undergo a significant decrease in their $T_1$. The quenching ratio $\eta=T_{1,\rm bare}/T_{1,\rm treated}$ is found to be $\eta\approx 7$ on average after the first treatment with 1 mM of Gd$^{3+}$ solution [Fig. \ref{Fig3}(b)]. From this value, we estimate a surface density of Gd$^{3+}$ spins $\sigma_{\rm Gd}\approx 4 \ {\rm nm}^{-2}$, corresponding to the detection of $\approx 1000$ spins for a $10$-nm ND~\cite{SI}. After the second treatment step with 10 mM of Gd$^{3+}$ solution, we obtain on average $\eta\approx 31$ corresponding $\sigma_{\rm Gd}\approx 70 \ {\rm nm}^{-2}$. The dispersion in the measured values for $\eta$ is attributed to a non uniform surface spin density $\sigma_{\rm Gd}$. In particular, NV defects located close to the diamond-substrate interface should be less affected by the Gd$^{3+}$ treatment. It is worth mentioning that a similar environment-induced quenching effect could be observed on the spin coherence time $T_2$ as well, with however much smaller quenching ratios since $T_{2}^{\rm bulk}\ll T_{1}^{\rm bulk}$~\cite{Steinert2012,Liam2012}. Furthermore, $T_1$-based sensing schemes do not require coherent manipulation of the NV electron spin with microwave pulses. \\ 
\begin{figure}[t]
\begin{center}
\includegraphics[width=0.48\textwidth]{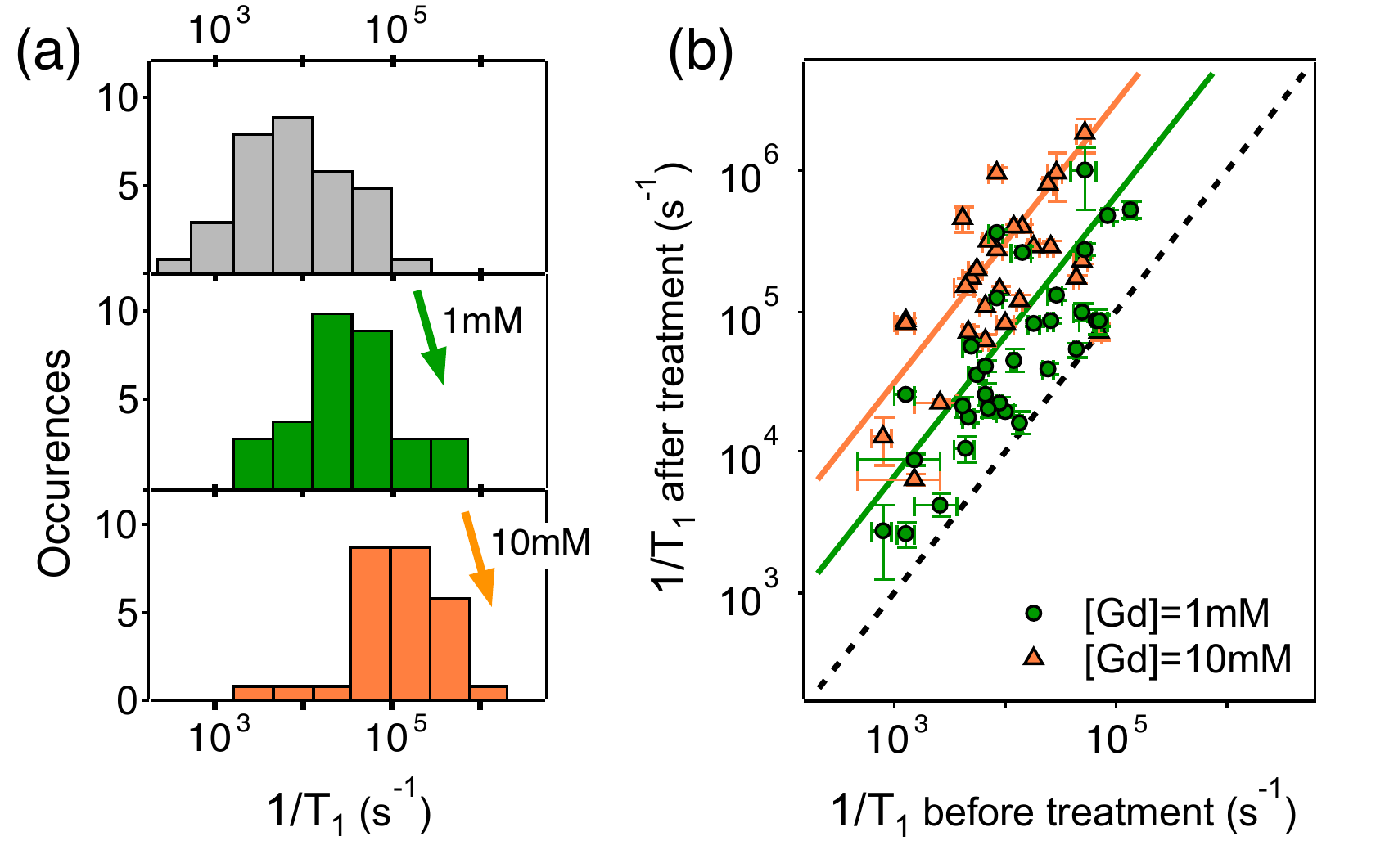}
\caption{(a)-Histograms of the $1/T_{1}$ relaxation rate obtained from a set of 33 single NV defects hosted in isolated NDs. The measurement is performed before any treatment (top panel), after adding 1 mM of the Gd$^{3+}$ solution (middle panel) and after further adding 10 mM of solution (bottom panel). (b)-Relaxation rate measured after the first (circles) and second (triangles) treatment step as a function of the rate of the bare ND. The solid lines are data fitting with linear functions whose slope indicates the average quenching ratio $\eta=T_{1,\rm bare}/T_{1,\rm treated}$. We obtain $\eta\approx 7$ (green line) and $\eta\approx 31 $ (orange line). A dashed line of slope $\eta=1 $ is plotted for reference.}
\label{Fig3}
\end{center}
\end{figure}
\indent By applying a few more treatment steps with the Gd$^{3+}$ solution, we then analysed the regime of strongly fluctuating magnetic environment, bringing $T_1$ in the sub-microsecond range. As shown in Fig.~\ref{Fig4}, $T_1$-quenching is accompanied by a significant reduction of the $T_1$ decay contrast, defined as $C_1^{\rm eff}={\rm max}[\mathcal{I}(\tau)]/\mathcal{I}(\infty)-1$. In the inset of Fig.~\ref{Fig4}, we plot $C_1^{\rm eff}$ as a function of $T_1$ together with the calculation based on a rate equation model that takes into account the full dynamics of the NV defect~\cite{Tetienne2012,SI}. The contrast reduction is mainly due to the overlap between $T_m$ and $T_1$ decays [see Eq.~(\ref{eqPL})]. In addition, when $T_1\sim T_m$, optical initialization in the $m_s=0$ state and spin state readout become less efficient, thus reducing further the contrast. \\ 
 \begin{figure}[t]
\begin{center}
\includegraphics[width=0.49\textwidth]{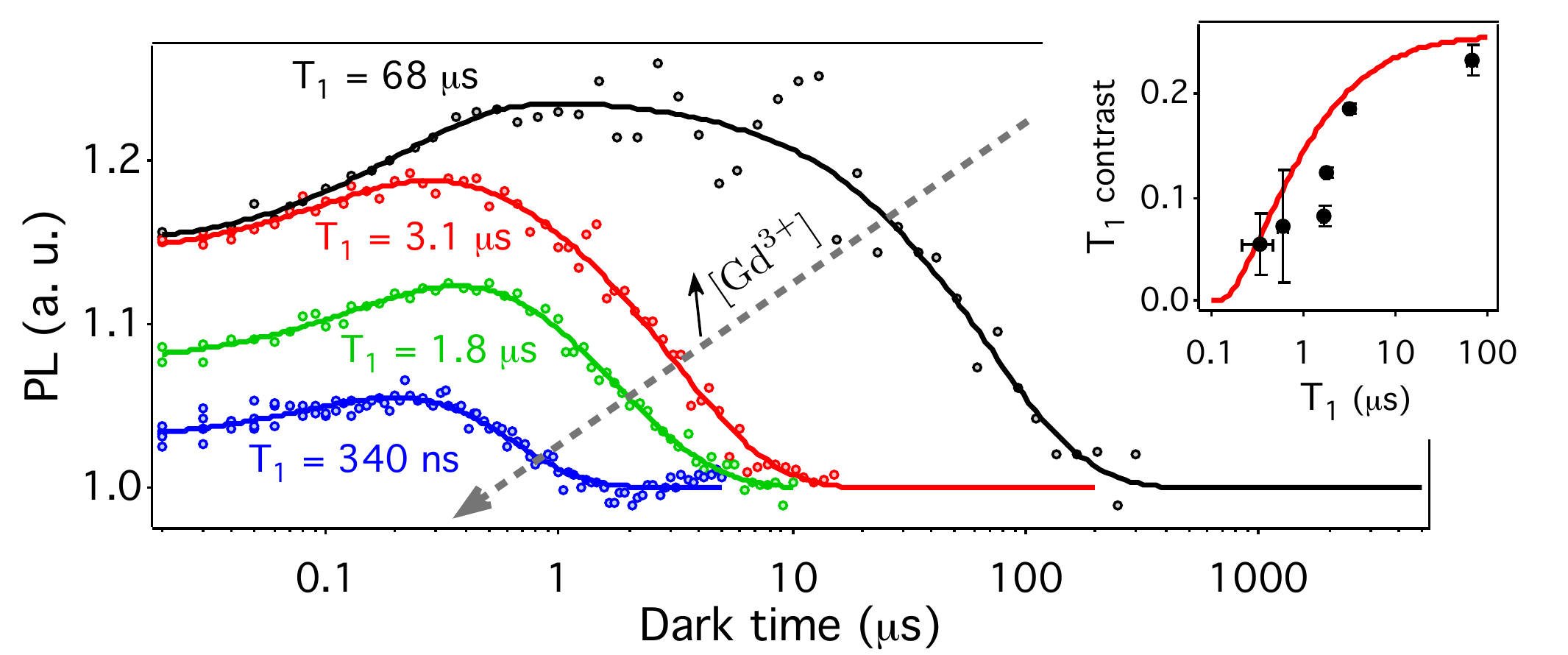}
\caption{$\mathcal{I}(\tau)$ relaxation curves measured after repeated treatments with the Gd$^{3+}$ solution. Solid lines are fit to Eq. (\ref{eqPL}). Inset: effective $T_1$ contrast $C_1^{\rm eff}$ as a function of $T_1$. The solid line is the result of the calculation (with no fit parameter) using a rate equation model with the parameters of NV centers in bulk diamond~\cite{SI}.}
\label{Fig4}
\end{center}
\end{figure}
\indent Based on our experimental results, we finally estimate the sensitivity of $T_1$ relaxometry to small changes in the magnetic environment. For that purpose, we consider an optimized single-$\tau$ measurement by fixing $\tau\sim T_{1}/2$, which converts a modification of spin relaxation into a change of the PL signal $\mathcal{I}(\tau)$ with optimal signal-to-noise ratio~\cite{Steinert2012,SI}. Assuming a photon shot noise limited signal, the smallest number of additional surface electronic spins $\delta{\cal N}_{\rm min}$ that can be detected by a single NV defect located at the center of a ND with size $d_0$ is given by
\begin{equation}
\delta{\cal N}_{\rm min}=\frac{1}{\mathcal{P}\sqrt{\Delta t}}d_0^4 f(\sigma) \ ,
\end{equation}
where $\mathcal{P}$ includes both the finite contrast of the $T_1$ relaxation signal and the rate of detected photons, $\Delta t$ is the integration time and $f(\sigma)$ is a slowly increasing function of the intrinsic density of surface spins $\sigma$~\cite{SI}. As expected, it is crucial to use NDs as small as possible in view of sensing magnetic noise from external spins. For a single NV defect hosted in a 10-nm ND with $\sigma=1$ nm$^{-2}$ -- {\it i.e.} corresponding to $T_1=6.3$ $\mu$s according to the above model [Fig.~\ref{Fig2}(b)] --,  a typical photon counting rate ${\cal R}=10^5$ s$^{-1}$ under cw optical illumination and a $T_1$ contrast $C_1=0.2$, we find $\delta{\cal N}_{\rm min} =14$ spins within $10$~s of integration. This result highlights that $T_1$ relaxometry with a single NV spin hosted in a ND is a promising resource to probe nanoscale magnetic field fluctuations with a sensitivity down to a few electron spins, within a time scale that is compatible with scanning probe techniques. Such probes might find important applications in life sciences, {\it e.g.} to image the spin density in biological samples with an unprecedented spatial resolution.

\indent The authors thank J. Lautru, S. Steinert and F. Ziem for experimental assistance and fruitful discussions. This work was supported by C'Nano Ile-de-France and the ANR projects D{\sc iamag}, A{\sc dvice} and Q{\sc invc}.

\begin{widetext}
\vspace{0.5cm}
\section{Supplementary Information} 

\subsection{Experimental setup}

All measurements were performed using a combined AFM/confocal microscopy setup operating under ambient conditions. A detailed description of the experimental setup can be found in Ref.~[\onlinecite{Rondin2012bis}]. Single NV defects hosted in diamond nanocrystals were optically excited with a laser operating at the wavelength $\lambda=532$ nm and their photoluminescence (PL) in the $650-800$~nm range was detected using an avalanche photodiode. Laser pulses were produced with an acousto-optical modulator (MT200-A0.5-VIS) with a characteristic rising time of $10$~ns. For all experiments, the optical pumping power was set at $1$~mW, corresponding to the saturation power of the NV defect radiative transition. 

\subsection{Derivation of the relaxation curve $\mathcal{I}(\tau)$}

We first consider the NV defect ground state as a {\it closed} three-level system composed of the three spin sublevels $m_s=0,\pm1$ with populations $n_{0,\pm1}$. The spin sublevels $m_s=0$ and $m_s=\pm1$ are coupled by two-way transition rates of strength $k_{01}$ [see Fig.~1-(a) of the main article]. After initialization into the $m_s=0$ spin sublevel with an optical pulse, the population dynamics are given by 
\begin{eqnarray} \label{eqPop1}
n_0(\tau)&=&\frac{1}{3}+\left[n_0(0)-\frac{1}{3}\right]e^{-\tau/T_1}\\
n_{\pm1}(\tau)&=&\frac{1}{3}\pm\left[\frac{n_{+1}(0)-n_{-1}(0)}{2}\right]e^{-\tau/(3T_1)}-\frac{1}{2}\left[n_0(0)-\frac{1}{3}\right]e^{-\tau/T_1} \ ,
\end{eqnarray}
where $T_1=[3k_{01}]^{-1}$ and $n_{i}(0)$ is the initial population of spin sublevel $m_s=i$.\\
\indent As indicated in the main text, the readout PL signal $\mathcal{I}(\tau)$ can be written as 
\begin{equation}
\label{S}
\mathcal{I}(\tau)=A_0n_0(\tau)+A_1[n_{+1}(\tau)+n_{-1}(\tau)] \ ,
\end{equation}
where $A_0$ and $A_1$ are the PL rates associated with spin states $m_s=0$ and $m_s=\pm 1$. Owing to spin-dependent PL of the NV defect, we consider $A_1<A_0$. Using Eq.~(\ref{eqPop1}), the signal $\mathcal{I}(\tau)$ can then be written as
\begin{equation} \label{eqPL1}
\mathcal{I}(\tau)=\mathcal{I}(\infty)\left[1+C_1 e^{-\tau/T_1}\right]
\end{equation}
\indent with 
\begin{eqnarray*} \label{eqPL2}
\mathcal{I}(\infty)&=&\frac{A_0+2A_1}{3} \ , \\
C_1&=&\frac{A_0-A_1}{A_0+2A_1}\left[3n_0(0)-1\right] \ . 
\end{eqnarray*}
\indent As expected, the readout contrast $C_1$ is proportional to the difference of spin-dependent PL rate ($A_0-A_1$), and to the amount of initial spin polarization into the $m_s=0$ spin sublevel.

We now take into account the lowest-lying singlet state of the NV defect, thereafter referred to as the metastable state with population $n_m$ [see Fig.~1-(a) of the main article]. This population decays towards $m_s=0$ with a rate $k_{m0}=\alpha k_m$ and towards $m_s=\pm1$ with a rate $k_{m1}=(1-\alpha)k_m/2$. The metastable decay time is therefore given by $T_m=k_m^{-1}=(k_{m0}+2k_{m1})^{-1}$. Typically $\alpha\sim 0.5$ and $k_m^{-1}=T_m\approx 200$ ns~[\onlinecite{Robledo2011bis,Tetienne2012bis}]. Within this four-level model, the populations after a time $\tau$ read
\begin{eqnarray*} \label{eqPop3}
&n_m(\tau)&=n_m(0)e^{-\tau/T_m} \ ,\\
&n_0(\tau)&=\frac{1}{3}+\left[n_0(0)-\frac{1}{3}\right]e^{-\tau/T_1}+n_m(0)\left[\alpha T_1- \frac{T_m}{3}\right]\frac{e^{-\tau/T_1}-e^{-\tau/T_m}}{T_1-T_m} \ , \\
&n_{\pm1}(\tau)&=\frac{1}{3}\pm\left[\frac{n_{+1}(0)-n_{-1}(0)}{2}\right]e^{-\tau/(3T_1)}-\frac{1}{2}\left[n_0(0)-\frac{1}{3}\right]e^{-\tau/T_1}\\
& &-\frac{n_m(0)}{2}\left[\frac{\alpha T_1-\frac{T_m}{3}}{T_1-T_m}e^{-\tau/T_1}+\frac{(1-\alpha)T_1-\frac{2T_m}{3}}{T_1-T_m}e^{-\tau/T_m}\right] \ .
\end{eqnarray*}
Using Eq.~(\ref{S}), the readout PL signal can then be written as
\begin{equation} \label{eqPL4}
\mathcal{I}(\tau)=\mathcal{I}(\infty)\left[1-C_m e^{-\tau/T_m}+C_1 e^{-\tau/T_1}\right] \ ,
\end{equation}
corresponding to Eq.~(1) of the main paper with
\begin{eqnarray} \label{eqPL5}
\mathcal{I}(\infty)&=&\frac{A_0+2A_1}{3} \ , \\ 
C_1&\approx&\frac{A_0-A_1}{A_0+2A_1}[3n_0(0)-1+n_m(0)] \ , \\
C_m&\approx&n_m(0) \ .
\end{eqnarray}   
Note that we have assumed $\alpha\approx 1/3$ in order to obtain a simple formula for $C_1$.

Although $C_1$ does not depend explicitly on $T_1$, the initial populations $n_0(0)$ and $n_m(0)$ as well as the PL rates $A_0$ and $A_1$ do, as soon as $1/T_1$ is not negligible compared to all other transition rates involved in the initialization process. Since $T_m$ is the longest decay time of the NV center's dynamics besides $T_1$, this means that $C_1$ is constant as long as $T_1\gg T_m$. However, when $T_1$ gets closer to $T_m$, the longitudinal spin relaxation prevents efficient spin initialization in $m_s=0$ through optical pumping, {\it i.e.} $n_0(0) \rightarrow 1/3$, as well as efficient spin state readout, {\it i.e.} $(A_0-A_1)$ vanishes. This explains why the contrast of $T_1$ decay strongly decreases when $T_1\sim T_m$ [see Fig.~4 of the main paper]. 

In order to gain further insight on this contrast reduction of $T_1$ relaxation curves, the NV defect dynamics was modeled using the seven-level model shown in Fig. \ref{FigS2}(a). A detailed description of this model can be found in Ref.~[\onlinecite{Tetienne2012bis}]. We used classical rate equations to calculate the time evolution of the populations. The optical pumping parameter $\beta$ was varied in time in order to simulate the exact sequence employed in the experiment [see inset in Fig. 1(c) of the main article]. We used $\beta=1$ when the laser is on and $\beta=0$ when the laser is off. The PL was numerically integrated over the first $300$~ns of the readout optical pulse, as in the experiment. Apart from $k_{01}=1/3T_1$ that was taken as a variable, the values of the other transition rates were taken from Ref.~[\onlinecite{Robledo2011bis}], namely $k_r=65$ $\mu$s$^{-1}$, $k_{0m}=11$ $\mu$s$^{-1}$, $k_{1m}=80$ $\mu$s$^{-1}$, $k_{m0}=3$ $\mu$s$^{-1}$ and $k_{m1}=1.3$ $\mu$s$^{-1}$. The dark time evolution $\tau$ was varied in order to simulate the relaxation curves $\mathcal{I}(\tau)$ for different values of $T_1$, as depicted in Fig. \ref{FigS2}(b). From these simulations, the effective contrast $C_1^{\rm eff}={\rm max}\left[\mathcal{I}(\tau)\right]/\mathcal{I}(\infty)-1$ was inferred as a function of $T_1$, leading to the solid line shown in the inset of Fig.~4 of the main article.

In Fig. \ref{FigS2}(c), this line is plotted again, together with the experimental data from a set of 6 single NV centers in distinct diamond nanocrystals, including the data shown in Fig.~4 of the main article (NV B6). The model captures correctly, on average, the observed behavior of contrast versus $T_1$. However, we note that the investigated NV defects all exhibit a different maximum contrast. This feature can be explained by variations of the transition rates for different NV defects in nanocrystals, which depend on the local strain, the presence of nearby defects, etc. In Fig. \ref{FigS2}(d) we plotted the $\mathcal{I}(\tau)$ relaxation curves measured for NV A3, where we observe the extreme case of a nearly flat response, from which the $T_1$ time could not be determined with a meaningful uncertainty.

\begin{figure}[t]
\begin{center}
\includegraphics[width=0.7\textwidth]{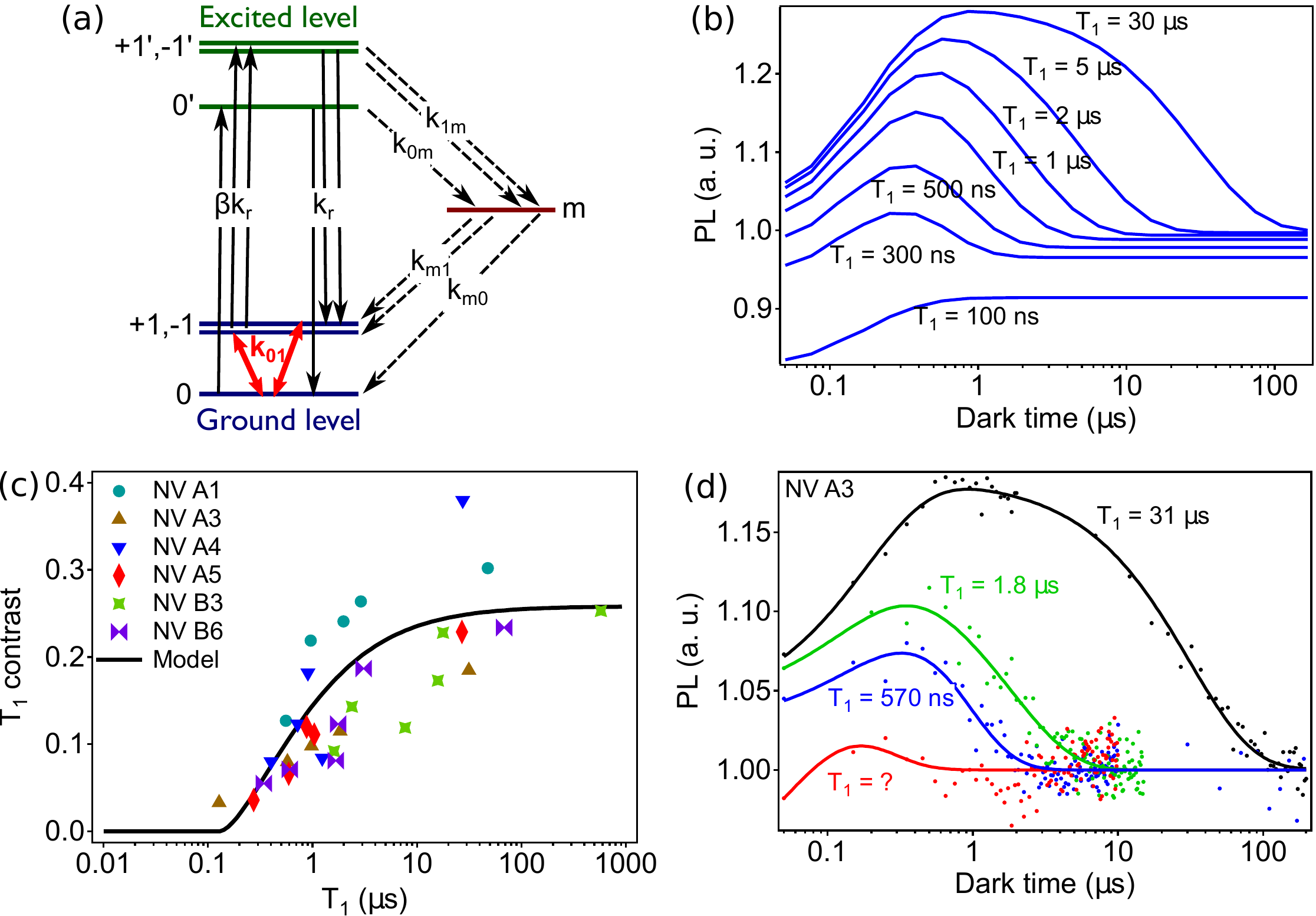}
\caption{(a)-Rate equation model used to simulate the $\mathcal{I}(\tau)$ relaxation curves. (b)-Simulated $\mathcal{I}(\tau)$ curves for various $T_1$ times, as described in the text. (c)-Effective $T_1$ contrast measured for various single NV defects as a function of $T_1$, which was tuned by adding Gd$^{3+}$ ions on the nanodiamond surface. The solid line is obtained from the simulated $\mathcal{I}(\tau)$ curves, according to the full model depicted in (a). (d)-Measured $\mathcal{I}(\tau)$ curves corresponding to NV A3. }
\label{FigS2}
\end{center}
\end{figure}

\subsection{Derivation of $T_1$ for a single NV defect interacting with a bath of surface spins}

Equation (3) of the main article gives the $1/T_1$ relaxation rate of the central NV spin placed in a fluctuating magnetic field $B(t)$ with zero mean, characterized by a variance $B_{\perp}^2$ (orthogonal to the NV defect quantization axis) and a correlation time $\tau_c$, {\it i.e.}
\begin{equation} \label{eqT12}
\frac{1}{T_{1}}=\frac{1}{T_{1}^{\rm bulk}} + 3\gamma_e^2 B_{\perp}^2 \frac{\tau_c}{1+\omega_0^2 \tau_c^2} \ ,  
\end{equation} 
where $\omega_0=2\pi D$ is the electron spin resonance (ESR) frequency of the NV defect.

In this section, we calculate the two quantities $B_{\perp}^2$ and $\tau_c$ for a bath of electronic spins distributed on the surface of the nanodiamond, modeled as a sphere of diameter $d_0$. The density of surface spins is denoted $\sigma$. 

\subsubsection{Variance of the transverse magnetic field}
\begin{figure}[t]
\begin{center}
\includegraphics[width=0.5\textwidth]{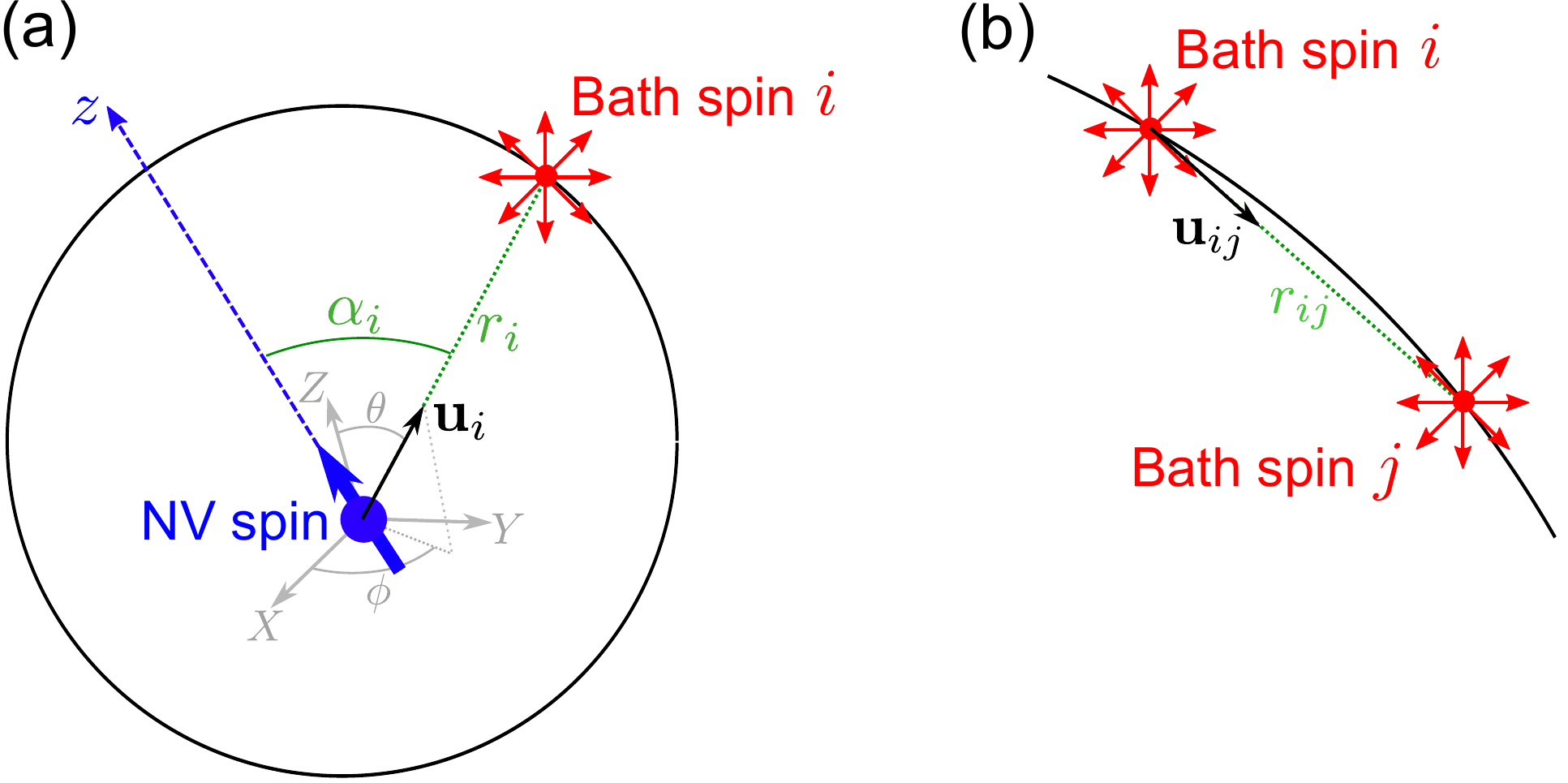}
\caption{(a) Notations for the derivation of the variance $B_{\perp}^2$ of the transverse magnetic field at the NV's location. The symmetry axis of the NV center is denoted $z$, while the $XYZ$ reference frame serves to define the spherical angles $(\theta,\phi)$ that describe the position of the bath spins on the sphere (vector ${\bf u}_{i}$ for spin $i$). The two frames $xyz$ and $XYZ$ can be chosen to be different for mathematical convenience (if the NV spin is not at the center of the sphere) when performing the integration. (b) Notations for the derivation of  the fluctuation rate $R_{\rm dip}$ of the spin bath caused by spin-spin interactions. }
\label{FigS1}
\end{center}
\end{figure}

The variance of the transverse magnetic field $B_{\perp}^2$ felt by the central spin can be obtained by summing over the contributions of all bath spins ${\bf S}_i$ according to $B_{\perp}^2=\sum_i B_{\perp,i}^2$. The dipolar field radiated by a spin ${\bf S}_i$ located at position ${\bf r}_i$ on the surface is given by 
\begin{equation} \label{eqB_1}
{\bf B}_i=\frac{\mu_0 \gamma_e \hbar}{4\pi r_{i}^3}\left[ {\bf S}_i - 3({\bf S}_i\cdot {\bf u}_{i}){\bf u}_{i}) \right] \ ,
\end{equation}  
where ${\bf u}_{i}={\bf r}_{i}/r_{i}$ [see notations in Fig. \ref{FigS1}(a)]. Tracing over a purely mixed state, described by a density matrix $\rho=\frac{1}{2S+1}\mathbbm{1}_{2S+1}$ where $\mathbbm{1}_{2S+1}$ is the identity matrix of size $2S+1$, we obtain
\begin{equation} \label{eqB_2}
B_{\perp,i}^2 = \langle B_{x,i}^2 \rangle + \langle B_{y,i}^2 \rangle = {\rm Tr}\lbrace\rho (B_{x,i}^2+B_{y,i}^2)\rbrace = \left(\frac{\mu_0 \gamma_e \hbar}{4\pi}\right)^2 C_S \frac{2+3\sin^2 \alpha_i}{r_i^6} \ ,
\end{equation}  
where $C_S=\frac{1}{2S+1}\sum_{m=-S}^S m^2=\frac{S(S+1)}{3}$ and $\alpha_i$ is the angle between ${\bf r}_i$ and the $z$-axis, which is the quantization axis of the central NV spin. For a bath of surface density $\sigma$, whose surface is described in spherical coordinates by $r=r(\theta,\phi)$ [Fig. \ref{FigS1}(a)], the summation over all the bath spins gives
\begin{equation} \label{eqB_3}
B_{\perp}^2 = \sum_i B_{\perp,i}^2= \left(\frac{\mu_0 \gamma_e \hbar}{4\pi}\right)^2 C_S\sigma \int_0^{2\pi} {\rm d}\phi \int_0^\pi {\rm d}\theta \sin \theta \frac{2+3\sin^2(\alpha(\theta,\phi))}{r(\theta,\phi)^4} \ .
\end{equation} 
For a spherical surface of diameter $d_0$ and a central spin located exactly at the center of the sphere, we have $r=d_0/2$ and $\alpha(\theta,\phi)=\theta$, which yields 
\begin{equation} \label{eqB_4}
B_{\perp}^2 =  \left(\frac{4\mu_0 \gamma_e \hbar}{\pi}\right)^2 \pi C_S \frac{\sigma}{d_0^4} \ .
\end{equation}
Considering $S=1/2$ surface spins, we find $B_{\perp}=26$ mT nm$^3\times\frac{\sigma^{1/2}}{d_0^2}$. For instance, $B_{\perp}=260$ $\mu$T-rms for a 10-nm nanocrystal with $\sigma=1$ nm$^{-2}$.  

If the central spin is offset by $\delta r$  along the $z$-direction, then $r(\theta,\phi)=\sqrt{(d_0/2)^2-\delta r^2 \sin^2 \theta}+\delta r \cos \theta$ and $\alpha(\theta,\phi)=\theta$ in Eq. (\ref{eqB_3}). An offset along the transverse direction, {\it e.g.} along the $x$-axis, can be taken into account using the same formula for $r(\theta,\phi)$ but a modified angle $\alpha(\theta,\phi)$ such that $\sin^2 (\alpha(\theta,\phi))=\Vert {\bf u}_r \times {\bf u}_x \Vert^2=\cos^2 \theta + \sin^2 \theta \sin^2 \phi$, instead of $\sin^2 (\alpha(\theta,\phi))=\Vert {\bf u}_r \times {\bf u}_z \Vert^2=\sin^2 \theta$. The latter case is the one considered to obtain the red dashed line in Fig.~2(b) of the main paper, where the NV defect is located 3 nm below the nanocrystal surface with its quantization axis ($z$) parallel to the surface.

\subsubsection{Correlation time of the bath}

The total fluctuation rate $R=1/\tau_c$ of the bath spins can be decomposed into two main contributions $R=R_{\rm dip}+R_{\rm vib}$ where $R_{\rm dip}$ is due to intra-bath dipolar coupling while $R_{\rm vib}$ is caused by intrinsic vibrational spin relaxation. 

An estimate for $R_{\rm dip}$ can be obtained by summing the dipolar interactions of a given spin ${\bf S}_i$ with all other spins ${\bf S}_j$ of the bath, according to $\hbar R_{\rm dip}=\sqrt{\sum_{j\neq i} \langle H_{ij}^2 \rangle}$ where $H_{ij}$ is the magnetic dipolar interaction   
\begin{equation} \label{eqR_1}
H_{ij}=\frac{\mu_0 \gamma_e^2 \hbar^2}{4\pi r_{ij}^3}\left[ {\bf S}_i\cdot{\bf S}_j - 3({\bf S}_i\cdot {\bf u}_{ij})({\bf S}_j\cdot {\bf u}_{ij}) \right] \ .
\end{equation}   
Here ${\bf r}_{ij}$ is the radius-vector between the two spins and ${\bf u}_{ij}={\bf r}_{ij}/r_{ij}$, as depicted in Figure~\ref{FigS1}(b). For purely mixed spin states, the density matrix describing the two-spin system is $\rho=\frac{1}{(2S+1)^2}\mathbbm{1}_{2S+1} \otimes \mathbbm{1}_{2S+1}$. The quantity $\langle H_{ij}^2 \rangle={\rm Tr}\lbrace\rho H_{ij}^2\rbrace$ is then given by 
\begin{equation} \label{eqR_2}
\langle H_{ij}^2 \rangle =  \left(\frac{\mu_0 \gamma_e^2 \hbar^2}{4\pi}\right)^2  \frac{6C_S^2}{r_{ij}^6} \ .
\end{equation}    
For a bath of surface spins with a surface density $\sigma$, we consider
\begin{equation} \label{eqR_3}
 \sum_{j\neq i}\frac{1}{r_{ij}^6} \approx \sigma\int_{r_{\rm min}}^{+\infty}\frac{2\pi r}{r^6}{\rm d}r \ ,
\end{equation}  
where $r_{\rm min}$ is the minimum allowed distance between two surface spins -- {\it e.g.} the lattice constant for a crystal. The final expression for $R_{\rm dip}$ reads
\begin{equation} \label{eqR_4}
\hbar R_{\rm dip} = \sqrt{\sum_{j\neq i} \langle H_{ij}^2 \rangle} = \frac{\mu_0 \gamma_e^2 \hbar^2 \sqrt{6} C_S}{4\pi} \sqrt{\frac{\pi}{2}}\frac{\sigma^{1/2}}{r_{\rm min}^{2}}. 
\end{equation}    
For a bath of electronic spins with $S=1/2$ and using $r_{\rm min}=0.15$ nm -- {\it i.e.} the nearest neighbour distance in diamond -- we find $R_{\rm dip}=11$ ns$^{-1}$ nm $\times\sigma^{1/2}$. Hence $R_{\rm dip}=11$~ns$^{-1}=2\pi \times 1.8$ GHz by using $\sigma=1$ nm$^{-2}$. 

Several EPR studies on various types of nanodiamonds have shown that the vibrational spin relaxation $R_{\rm vib}$ of the SPCs is around 1~ns$^{-1}$~[\onlinecite{Panich2011bis,Dubois2009bis}] or less~[\onlinecite{Casabianca2011bis,Orlinskii2011bis}]. These values are much smaller than $R_{\rm dip}$, as calculated above for a typical surface spin density $\sigma=1$ nm$^{-2}$. Therefore, we can neglect the vibrational contribution to the overall fluctuation rate, so that $R\approx R_{\rm dip}$. 

\subsubsection{$T_{1}$ relaxation as a function of the nanodiamond size}

Replacing Eq. (\ref{eqB_4}) in Eq. (\ref{eqT12}), {\it i.e.} for a NV spin located at the center of a ND of diameter $d_0$, the longitudinal spin relaxation rate writes
\begin{equation} \label{eqT1_3}
\frac{1}{T_{1}}=\frac{1}{T_{1}^{\rm bulk}}+ \left(\frac{48\mu_0^2\gamma_e^4\hbar^2C_s}{\pi d_0^4} \right) \left( \frac{\sigma R(\sigma)}{\omega_0^2+R(\sigma)^2} \right) \ ,
\end{equation} 
where $R(\sigma)=1/\tau_c\approx R_{\rm dip}(\sigma)$ is given by Eq. (\ref{eqR_4}). It can be seen that the second term in Eq. (\ref{eqT1_3}), which dominates in NDs, scales as $d_0^{-4}$. This dependence, which simply stems from the $d_0^{-6}$ of the spin-spin interaction integrated over a surface, is responsible for the variation of $T_1$ over several orders of magnitude when the size of the ND decreases. 
\begin{figure}[t]
\begin{center}
\includegraphics[width=0.60\textwidth]{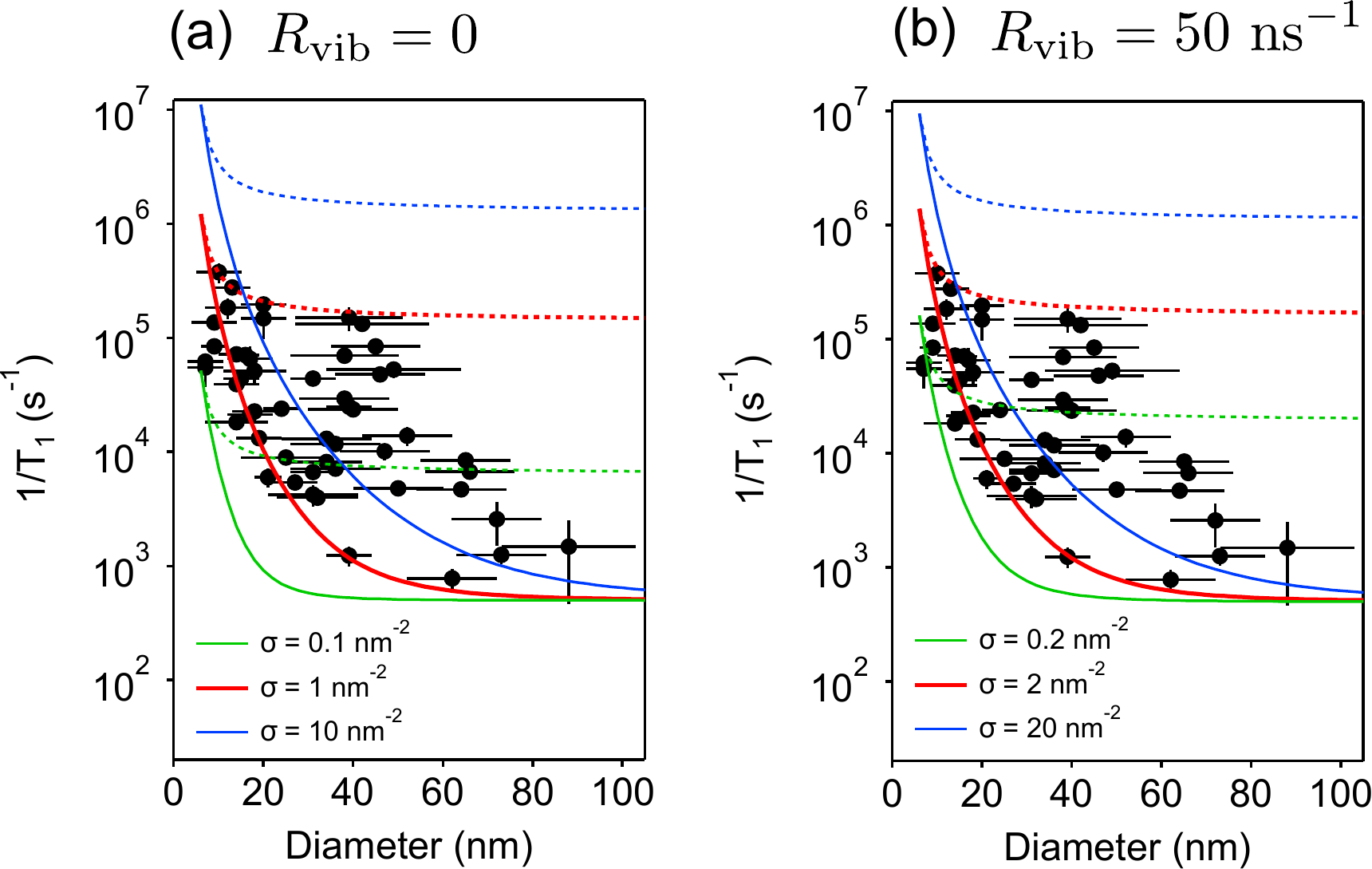}
\caption{ (a,b) Longitudinal spin relaxation rate $1/T_1$ of the NV defect electron spin as a function of the ND diameter. The markers are experimental data while the lines are the results of the calculation for a NV spin located at the center of the sphere (solid line) and $3$ nm below the surface (dotted line), for several values of the surface spin density $\sigma$. We assumed $ T_{1}^{\rm bulk}=2$ ms. We considered a vibrational contribution to the spin bath fluctuation rate $R_{\rm vib}=0$ in (a), as in the main article, and $R_{\rm vib}=50$~ns$^{-1}$ in (b).}
\label{FigS4}
\end{center}
\end{figure}

In Fig. \ref{FigS4}(a), we plot again the $1/T_1$ vs. $d_0$ data points, together with the calculation using the above formulas. The calculation is performed for the two extreme cases where the NV spin is located either at the center of the sphere [{\it best case}, given by Eq. (\ref{eqT1_3})] or $3$~nm away from the surface with the NV axis parallel to the surface ({\it worst case}). The result of the calculation is shown for three different surface densities $\sigma$. The value $\sigma\approx 1$~nm$^{-2}$ is the one that best `fits' the experimental results, if the criterion is that most data points must lie in between the two extreme case theoretical curves. Figure \ref{FigS4}(b) is identical to Fig. \ref{FigS4}(a) but here we assumed a non-zero value for the vibrational contribution to the correlation time, namely $R_{\rm vib}=50$~ns$^{-1}$, which is a typical value for electron spins in paramagnetic complexes~[\onlinecite{Kruk2004bis}], but is probably an extreme case for our nanodiamond SPCs. Now the experimental results are best `fitted' with a surface density $\sigma\approx 2$~nm$^{-2}$. This shows that the outcome of the model is not very sensitive to the choice of $R_{\rm vib}$ for reasonable values.

\subsection{Estimation of the number of gadolinium spins}

In this section we estimate the number of gadolinium spins that were added to the ND surface from the experimentally measured quenching ratio $\eta=T_{1,\rm bare}/T_{1,\rm treated}$ [see Fig. 3 of the main text]. Here $T_{1,\rm treated}$ (resp. $T_{1,\rm bare}$) denotes the relaxation time of the NV defect after (resp. before) adding Gd$^{3+}$ ions. Since these experiments were performed in the limit $\eta\gg 1$, one can consider that $T_{1,\rm treated}$ is only due to a bath of $S'=7/2$ Gd$^{3+}$ spins with a density $\sigma_{\rm Gd}$, producing at the NV location a fluctuating magnetic field characterized by a variance $B_{\perp}'^2$ and a correlation time $\tau_c'$. The spin relaxation rate after adding Gd$^{3+}$ ions can therefore be written    
\begin{equation} \label{eqGd1}
\frac{1}{T_{1,\rm treated}}\approx 3\gamma_e^2 B_{\perp}'^2 \frac{\tau_c'}{1+\omega_0^2 \tau_c'^2} \ .  
\end{equation} 
Neglecting $1/T_1^{\rm bulk}$ and considering only the intra-bath dipolar contribution to $\tau_c'$, one obtains from Eqs. (\ref{eqB_4}) and (\ref{eqR_4}) that
\begin{eqnarray} \label{eqGd2}
\eta & \approx & \frac{C_{S'}\sigma_{\rm Gd}}{C_{S}\sigma}\times\frac{C_{S}\sigma^{1/2}}{C_{S'}\sigma_{\rm Gd}^{1/2}}\times\frac{1+\omega_0^2 \tau_c^2}{1+\omega_0^2 \tau_c'^2} \\
& \approx & 3.64 \left( \frac{\sigma_{\rm Gd}}{\sigma} \right) ^{1/2}.
\end{eqnarray}  
In the latter expression, we assumed $\sigma=1$~nm$^{-2}$, {\it i.e.} $1/\tau_c=11 \ {\rm ns}^{-1}$ and $1/\tau_c'\gg \omega_0$ as justified below. Within this framework, $\eta$ depends neither on the ND size nor on its exact shape, in agreement with the experimental data. By adding 1 mM of Gd$^{3+}$ solution, an averaged quenching ratio $\eta=7$ is obtained [see Fig. 3 of the main text]. From this value, we estimate a surface density of gadolinium spins $\sigma_{\rm Gd}\approx 4\sigma\approx 4$~nm$^{-2}$. For a $10$-nm size ND, this corresponds to the detection of $\approx 1000$ gadolinium spins. After adding 10 mM of solution, we find $\eta=31$ on average, corresponding to $\sigma_{\rm Gd}\approx 70\sigma\approx 70$~nm$^{-2}$.

We note that $\sigma_{\rm Gd}=4$~nm$^{-2}$ yields $R_{\rm dip}'\approx500$~ns$^{-1}$, which is indeed much larger than the vibrational contribution (typically $R_{\rm vib}'\sim50$~ns$^{-1}$ [\onlinecite{Kruk2004bis}]) and than $\omega_0=18$~ns$^{-1}$. This validates the assumptions made to obtain Eq.~(25), {\it i.e.} that only the intra-bath dipolar term contributes to $\tau_c'$ and $1/\tau_c'\gg \omega_0$.

\subsection{Sensitivity estimation}

In this section we derive the sensitivity of $T_1$ relaxometry to a change in the number ${\cal N}_e$ of external electronic spins on the surface, hence to a change in the relaxation rate $\Gamma_1=1/T_1$ of the NV spin. We consider a single-$\tau$ detection scheme~[\onlinecite{Steinert2012bis}] that consists in measuring the PL signal $\mathcal{I}(\tau)$ after a fixed dark time $\tau \gg T_m$, as shown in Figure~\ref{FigS3}. The number of photons detected while repeating this sequence during a total acquisition time $\Delta t$ reads
\begin{equation} \label{eqS1}
N(\Gamma_1)\approx \frac{{\cal R}T_{\rm int}}{\tau}\Delta t\left[1+C_1 e^{-\Gamma_1 \tau}\right] \ ,
\end{equation} 
where ${\cal R}$ is the photon counting rate under cw optical illumination, $T_{\rm int}$ is the width of the integration window (300 ns in our experiments), and $C_1\ll1$ is the contrast. An infinitesimal change $\delta\Gamma_1$ of the relaxation rate is converted into a modification $\delta N_{\rm signal}$ of the number of detected photons given by
\begin{equation} \label{eqS2}
\delta N_{\rm signal}\approx  \delta\Gamma_1{\cal R}T_{\rm int}\Delta tC_1 e^{-\Gamma_1 \tau} \ .
\end{equation} 
Assuming a photon shot noise limited signal $\delta N_{\rm noise}=\sqrt{N(\Gamma_1)}$ and $C_1\ll1$, the signal to noise ratio (SNR) reads 
\begin{equation} \label{eqS4}
{\rm SNR}=\frac{\delta N_{\rm signal}}{\delta N_{\rm noise}} = \delta\Gamma_1 \sqrt{{\cal R}T_{\rm int}\tau\Delta t}C_1 e^{-\Gamma_1 \tau} \ .
\end{equation} 
The maximal signal-to-noise ratio SNR$_{m}$ is obtained for $\tau=T_1/2$ and is given by
\begin{equation} \label{eqS5}
{\rm SNR}_{m}=\frac{\delta\Gamma_1}{\sqrt{\Gamma_1}} C_1\sqrt{\frac{{\cal R}T_{\rm int}\Delta t}{2e}} \ .
\end{equation} 
For a net change $\delta\Gamma_1$, the ${\rm SNR}$ is therefore improved when $\Gamma_1$ decreases. However $\delta\Gamma_1$ and $\Gamma_1$ are not independent quantities, as we shall see below. 

\begin{figure}[t]
\begin{center}  
\includegraphics[width=0.5\textwidth]{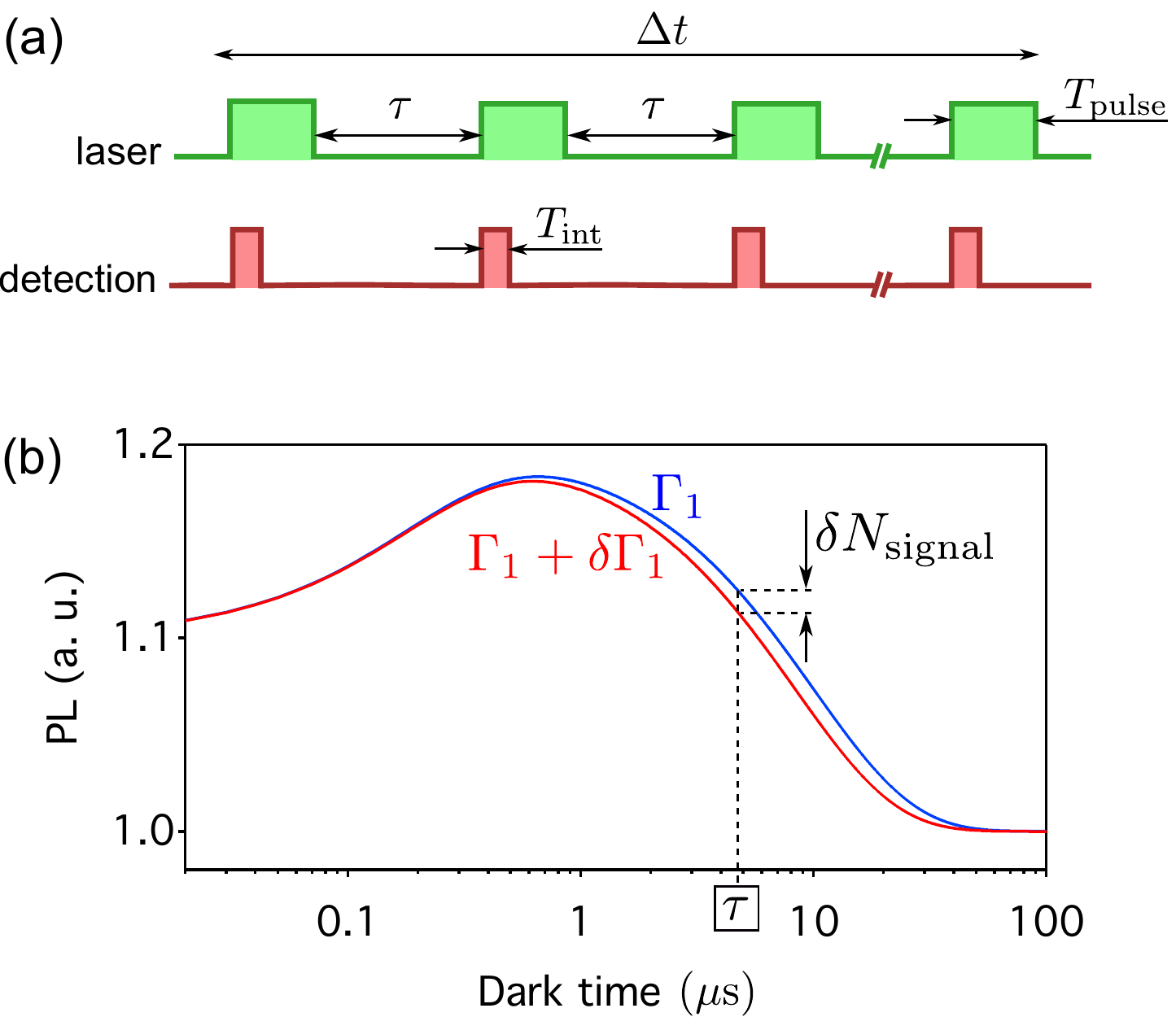}
\caption{(a)-Sequence used in the single-$\tau$ detection scheme. In the derivation of the sensitivity, the repetition period is taken to be equal to the dark time $\tau$, {\it i.e.} we neglect the duration $T_{\rm pulse}$ of the laser pulse, which can be as short as 1 $\mu$s in practice. (b)-Typical $\mathcal{I}(\tau)$ curve calculated with a spin relaxation rate $\Gamma_1=10^5$ s$^{-1}$ (blue curve). The red curve illustrates the effect of an increase of the relaxation rate by $\delta\Gamma_1=0.2\Gamma_1$. The single-$\tau$ detection scheme enables to probe the PL difference for a given dark time $\tau$, which is denoted $\delta N_{\rm signal}$ in the text and is related to the change $\delta\Gamma_1$.}
\label{FigS3}    
\end{center}
\end{figure}
In the following, we consider a single NV spin located at the center of a spherical diamond nanocrystal of diameter $d_0$. Using Eqs.~(\ref{eqR_4})-(\ref{eqT1_3}) and neglecting $1/T_1^{\rm bulk}$, the relaxation rate due to a bath of electronic spins $S=1/2$ with surface density $\sigma$ can be expressed as
\begin{equation} \label{eqS6}
\Gamma_1=\frac{A}{d_0^4} \frac{\sigma^{3/2}}{\sigma+B} \ ,
\end{equation} 
with $A=5.75\times 10^9$ s$^{-1}$nm$^5$ and $B=2.64$ nm$^{-2}$. We seek to calculate the effect of an infinitesimal increase $\delta\sigma$ of the surface spin density. The resulting modification of the relaxation rate $\delta\Gamma_1$ is given by 
\begin{equation} \label{eqS7}
\frac{\delta\Gamma_1}{\Gamma_1} = \frac{\delta\sigma}{\sigma}  \left(\frac{1}{2}+\frac{B}{B+\sigma}   \right).
\end{equation} 
Inserting Eqs. (\ref{eqS6}) and (\ref{eqS7}) into Eq.~(\ref{eqS5}), we finally obtain
\begin{equation} \label{eqS8}
{\rm SNR}_m=\delta\sigma \times \frac{C_1}{2 d_0^2} \sqrt{\frac{A{\cal R}T_{\rm int}\Delta t}{2e}} \frac{\sigma+3B}{\sigma^{1/4}(\sigma+B)^{3/2}} \ .
\end{equation} 

In terms of increase in the number of surface spins $\delta{\cal N}_e=\delta\sigma\pi d_0^2$, the SNR writes 
\begin{equation} \label{eqS9}
{\rm SNR}_m=\delta{\cal N}_e\times \frac{C_1}{2\pi d_0^4} \sqrt{\frac{A{\cal R}T_{\rm int}\Delta t}{2e}} \frac{\sigma+3B}{\sigma^{1/4}(\sigma+B)^{3/2}}.
\end{equation} 
Owing to the $d_0^{-4}$ dependence, it is crucial to use a nanodiamond as small as possible in order to detect a net change $\delta{\cal N}_e$. The smallest number of spins that can be detected is obtained for a signal to noise ratio of one, and reads
\begin{equation} \label{eqS9}
\delta{\cal N}_{e,min}=\frac{1}{\mathcal{P}\sqrt{\Delta t}}d_0^4 f(\sigma) \ ,
\end{equation} 
where $\mathcal{P}=\frac{C_1}{2\pi} \sqrt{\frac{A{\cal R}T_{\rm int}}{2e}} $ and $f(\sigma)=\frac{\sigma^{1/4}(\sigma+B)^{3/2}}{\sigma+3B}$ is a slowly fluctuating function of $\sigma$. 
For a single NV spin located at the center of a 10-nm diameter nanocrystal with an intrinsic spin density $\sigma=1$ nm$^{-2}$, a typical photon counting rate ${\cal R}=10^5$ s$^{-1}$ and a contrast $C_1=0.2$, we find $\delta{\cal N}_{e,min}\approx 14$ spins within 10 s of integration. We emphasize that although the above calculation assumes the $\delta{\cal N}_e$ additional spins to be distributed all over the nanocrystal's surface, we expect our conclusions, especially the distance dependence, to be true even if the $\delta{\cal N}_e$ spins are localized in a point-like spot, {\it e.g.} in a molecule that would be approached close to the nanocrystal's surface.

\end{widetext}


\begin{thebibliography}{50}
\bibitem{Note}
In contrast with NV defects in bulk diamond, $T_1$ could thus not be increased by lowering the temperature
\end{thebibliography}
\end{document}